%% file: paper.tex
\documentclass[sigconf,screen,dvipsnames]{acmart}
\usepackage{balance}

\usepackage{wrapfig}

\usepackage{footnote}
\makesavenoteenv{tabular}
\makesavenoteenv{table}
\usepackage{multirow}
\usepackage{xspace}

\usepackage{t4phonet}
\usepackage{pifont}

\newcommand{\tool}{{\sc ReproDroid}\xspace}

\usepackage{xcolor,colortbl}
\usepackage{listings}
\newcommand{\noop}[1]{}

\usepackage[disable,prependcaption,textsize=footnotesize,linecolor=OliveGreen,backgroundcolor=OliveGreen!25,bordercolor=OliveGreen]{todonotes}
\usepackage{letltxmacro}
\LetLtxMacro{\oldtodo}{\todo}
\renewcommand{\todo}[3][]{\oldtodo[#1]{\textbf{[#2]} #3}}

\usepackage{booktabs} % For formal tables

\setcopyright{acmlicensed}
\acmPrice{15.00}
\acmDOI{10.1145/3236024.3236029}
\acmYear{2018}
\copyrightyear{2018}
\acmISBN{978-1-4503-5573-5/18/11}
\acmConference[ESEC/FSE '18]{Proceedings of the 26th ACM Joint European Software Engineering Conference and Symposium on the Foundations of Software Engineering}{November 4--9, 2018}{Lake Buena Vista, FL, USA}
\acmBooktitle{Proceedings of the 26th ACM Joint European Software Engineering Conference and Symposium on the Foundations of Software Engineering (ESEC/FSE '18), November 4--9, 2018, Lake Buena Vista, FL, USA}

\begin{document}
%\title{Competitive Benchmarking: Do Android App Analysis Tools keep their Promises?}
\title{Do Android Taint Analysis Tools Keep Their Promises?}
%\titlenote{Produces the permission block, and copyright information}
%\subtitle{A Reproducibility Study} 
%\subtitle{Fair and automated benchmarking of Android App Analysis tools}
%\subtitlenote{The full version of the author's guide is available as texttt{acmart.pdf} document}

\author{Felix Pauck}
\affiliation{%
	\institution{University Paderborn}
	\streetaddress{Warburger Str. 100}
	\city{Paderborn}
	\country{Germany}
	\postcode{33098}
}
\email{fpauck@mail.uni-paderborn.de}

\author{Eric Bodden}
\affiliation{%
	\institution{University Paderborn}
	\streetaddress{Warburger Str. 100}
	\city{Paderborn}
	\country{Germany}
	\postcode{33098}
}
\email{eric.bodden@uni-paderborn.de}

\author{Heike Wehrheim}
\affiliation{%
	\institution{University Paderborn}
	\streetaddress{Warburger Str. 100}
	\city{Paderborn}
	\country{Germany}
	\postcode{33098}
}
\email{wehrheim@uni-paderborn.de}

\renewcommand{\shortauthors}{Felix Pauck, Eric Bodden, and Heike Wehrheim}

% Macros
\newcommand{\AQL}{{\sc AQL}\xspace}
\newcommand{\BREW}{{\sc Brew}\xspace}
\newcommand{\FlowDroid}{{\sc FlowDroid}\xspace}
\newcommand{\DroidBench}{{\sc DroidBench}\xspace}
\newcommand{\DIALDroid}{{\sc DIALDroid}\xspace}
\newcommand{\DidFail}{{\sc DidFail}\xspace}
\newcommand{\IccTA}{{\sc IccTA}\xspace}
\newcommand{\SuSi}{{\sc SuSi}\xspace}
\newcommand{\DroidSafe}{{\sc DroidSafe}\xspace}
\newcommand{\ApkCombiner}{{\sc ApkCombiner}\xspace}
\newcommand{\Amandroid}{{\sc Amandroid}\xspace}
\newcommand{\ICCBench}{{\sc ICC-Bench}\xspace}
\newcommand{\DIALDroidBench}{{\sc DIALDroid-Bench}\xspace}

\begin{abstract}
\input{content/abstract}
\end{abstract}




\maketitle

\bibliographystyle{ACM-Reference-Format}

\input{content/content}

\end{document}

%% file: content/abstract.tex
In recent years, researchers have developed a number of tools to conduct taint analysis of Android applications.
While all the respective papers aim at providing a thorough empirical evaluation, 
comparability is hindered by varying or unclear evaluation targets. 
%by the evaluation targets often differ from one publication to another. 
Sometimes, the apps used for evaluation are not precisely described.
In other cases, authors use an established benchmark but cover it only partially.
In yet other cases, the evaluations differ in terms of the data leaks searched for, or lack a ground truth to compare against.
All those limitations make it impossible to truly compare the tools based on those published evaluations. 
%All those reasons make it impossible to truly compare one tool to another, based on published evaluations.

We thus present \tool, a framework allowing the accurate comparison of Android taint analysis tools.
\tool supports researchers in inferring the ground truth for data leaks in apps, in automatically applying tools to benchmarks, and in evaluating the obtained results.
We use \tool to comparatively evaluate on equal grounds the six prominent taint analysis tools \Amandroid, \DIALDroid, \DidFail, \DroidSafe, \FlowDroid and \IccTA.
The results are largely positive although four tools violate some promises concerning features and accuracy.
Finally, we contribute to the area of unbiased benchmarking with a new and improved version of the open test suite \textsc{DroidBench}. 

%FlowDroid, DidFail, DroidSafe, IccTA and Amandroid represent some well-known examples of Android app analysis tools that aim at finding security issues such as privacy leaks.
%There exist many more and the number is steadily increasing.
%Each paper describing such a tool comes with a similar claim which asserts that the described tool is the best tool available in order to answer a certain set of analysis questions.
%So which tool is the overall best?
%Does such a tools exist?
%And do all these tools keep their promises?
%This paper deals with finding answers to these questions.
%To do so, we propose a toolchain that enables us to run a fair, automated and comparable benchmark of Android app analysis tools.
%For 6 well-known tools and 3 frequently used benchmarks, we execute this toolchain.
%Thereby, we turn out currently existing issues in these tools and benchmarks and highlight how to overcome these issues.
%Finally, we will also contribute an improved version of the most used benchmark, DroidBench.

%% file: content/content.tex
\section{Introduction}\label{sec:introduction}
\input{content/introduction}

\section{Background}\label{sec:background}
\input{content/background}

\section{Approach}\label{sec:approach}
\input{content/approach}

\section{Experimental setup}\label{sec:evaluation}
\input{content/evaluation}

\section{Experimental results}\label{sec:results}
\input{content/results}

\section{Related Work}\label{sec:related_work}
\input{content/relatedwork}

\section{Conclusion}\label{sec:conclusion}
\input{content/conclusion}

\begin{acks}
This work was partially supported by the German Research Foundation (DFG) within the Collaborative Research Centre ``On-The-Fly Computing'' (SFB 901).
\end{acks}

\newpage
\balance
\bibliography{content/references}

% TODO: Remove following
\newpage
\listoftodos{}

%% file: content/introduction.tex
With smartphones becoming  a part of everyone's daily life, users frequently downloading new apps to their phones and even performing security critical applications like banking, or the coordination of medical treatments, the probability of an undesired leak of private data increases steadily.
Such data leaks may arise form coding mistakes but also may be the result of malicious attacks. 
According to Gartner~\cite{gartner}, currently 86\% of all mobile phones 
use Android as operating system. 
%the most wide-spread operating system for mobile devices is Android. 
%86\% of all mobile devices such as smartphones, -watches or tablets use Android~\cite{gartner}.
In the past, researchers have proposed a number of tools to detect data leaks in Android applications.
The tools employ various analysis techniques, ranging from static \cite{DBLP:conf/ccs/WeiROR14, DBLP:conf/sigsoft/FengDAA14, DBLP:conf/ccs/BosuLYW17, DBLP:conf/pldi/KlieberFBJB14, DBLP:conf/ndss/GordonKPGNR15, DBLP:conf/pldi/ArztRFBBKTOM14, DBLP:conf/icse/0029BBKTARBOM15, scandroid, DBLP:conf/wisec/CuiWHXZY15} over dynamic \cite{DBLP:conf/osdi/EnckGCCJMS10, DBLP:conf/iscc/ZhangWLGC16} and hybrid \cite{DBLP:conf/sac/AhmadCCB17} analyses to methods built on logical reasoning \cite{DBLP:conf/eurosp/CalzavaraGM16}. 
Static analysis tools often employ a form of {\em taint} analysis: 
sensitive data (i.e., data from 
specific private sources) is tainted and then statically tracked through the application's data flow and sometimes control flow. 
Whenever tainted data reaches a pre-defined public sink, the taint analysis tool reports a privacy leak. 

Since several decades, static analysis is known to be undecidable~\cite{rice1953classes}. Undecidability forces static taint analysis tools to approximate the data flows they compute. 
In case of an over-approximation, the taint analysis might report spurious warnings, so-called \emph{false positives}, while in the case of under-approximations it may miss actual data flows, resulting in \emph{false negatives}.
While all static taint analysis tools naturally share the idea of tracking taints, each tool has its own strengths and weaknesses.
For instance, some but not all tools have good support for handling component lifecycles, callbacks or inter-component communication (ICC).
Further, the tools provide different levels of precision by supporting (or not) an object, field or context-sensitive analysis~\cite{DBLP:conf/popl/SmaragdakisBL11}.
Due to these various features, both researchers and practitioners wonder which tool is the optimal choice in which application context---a question that can only be answered through a comparative evaluation.

%Most users only see the handy aspects of such devices and do not hesitate to perform security sensitive operations on it.
%For example, using a banking app to transfer money is not a rarely performed action anymore.
%Taking photos and sharing them over the Internet has become a banality.
%However, who tells us that the money I transfer or the picture I take arrives at its intended destination?
%Who guarantees us that the picture I sent is not send anywhere else simultaneously?
%Android app analysis tools represent one fitting answer to such questions.
%For instance, there exist many tools that determine if sensitive information is intentionally or unintentionally leaked to the outside.
%To do so, so-called taint analyses are performed.

Many papers proposing taint-analysis tools evaluate those tools using benchmarks such as the open test suites \textsc{DroidBench}~\cite{DBLP:conf/pldi/ArztRFBBKTOM14} or \textsc{ICC-Bench}~\cite{DBLP:conf/ccs/WeiROR14}.
These so-called {\em micro-benchmarks} consist of artifical mini-apps developed for benchmarking purposes only.
Each app or a predefined combination of multiple apps represents one benchmark case. 
Benchmark cases contain intentionally encoded data leaks, e.g.\ flows of data from a statement accessing the device's serial number \texttt{getDeviceId()} to a statement sending SMS \texttt{sendTextMessage(...)}. 
While the usage of such common benchmark suites seems to provide a solid basis for an unbiased and systematic comparison, it suffers from some fundamental drawbacks:
For instance, the micro-benchmarks mostly lack information about the exact data leaks contained in each test case, i.e., the so-called {\em ground truth}.
Instead comparisons take place on the grounds of just {\em counting} the number of data leaks found and comparing it against the one given for the benchmark case.
This hides incorrect leak detections in cases where these numbers match up coincidentally.
As our experiments confirm, this problem exists not just in theory but has impaired past evaluations.

The comparison gets even less systematic when moving to {\em real-world apps}, i.e., apps that can be downloaded from an app market such as Google's Play store.\footnote{Google Play store~\url{https://play.google.com/store/apps}}
The inclusion of such apps in experiments is indispensable when it comes to evaluating the tools for scalability. 
Yet, in most recent works reporting the evaluation of Android taint analysis tools it is quite unclear which apps exactly have been used in the respective evaluation. To give one example of many, Li et al.\ state "We randomly selected 50 apps from our Googleplay set for our study." without naming them~\cite{DBLP:conf/icse/0029BBKTARBOM15}.
An additional related problem is that it is unclear which exact data leaks are to be found and have been found. 
Since for real-world apps not even the number of actual data leaks is known (let alone their exact data flows), the evaluations simply report the number of leaks found, without being able to assess which fraction of the tools' warnings might be false, and how many leaks are missed.
In fact, \emph{this way of measuring success rewards rather than penalizes false positives}, as higher numbers are frequently seen as better results.
In summary, tool benchmarks frequently lack confirmability and reproducibility.

The goal of this work is to remedy this unsatisfactory situation by enabling a reproducible, fair and unbiased comparison.
Specifically, we present \tool, a framework allowing for the  accurate and systematic benchmarking of Android taint analysis tools. 
%To overcome these issues, we present how to improve benchmarks and propose an approach that allows us to execute benchmarks and evaluate/compare benchmark results in a fair and automated way.
The approach comprises the following original contributions.
First, we introduce the \emph{Android App Analysis Query Language (\AQL)} which is used to precisely define analysis questions and answers. 
In questioning, the usage of \AQL\ allows us to run all examined tools on the same target, i.e., inspect the existence of the same flow of data.
On the answer side, \AQL\ acts as a standardized language for describing the flows found. 
Second, the associated \emph{\AQL-System} delegates analysis questions to appropriate tools by matching the question subject against tool capabilities, and converts the produced answers into the \AQL~format. 
Through those unique features, the \AQL-System supports the completely automatic benchmarking of tools. 
Third, we present the \emph{Benchmark Refinement and Execution Wizard} (\BREW), which helps one to refine, execute and evaluate precisely formulated benchmarks.
\BREW\ allows us and others to more easily determine the ground truth of data leaks for test apps. 

As another major contribution of this work, we use \tool to carry out a {\em reproducible} comparison of some of the most prominent taint analysis tools for Android apps: \Amandroid, \DIALDroid, \DidFail, \DroidSafe, \FlowDroid and \IccTA, on 265 apps (235 micro-benchmarks and 30 real-world apps). 
We find that the experiments reproduce most but not all results of previously published evaluations. 
To further contribute to the area of systematic benchmarking, we provide all benchmarks, now \emph{precisely} defined, within a new version of the open test suite \DroidBench.\footnote{\url{https://FoelliX.github.io/ReproDroid/\#droidbench}}%\footnote{Our pull request for \DroidBench is complete but, to not jeopardize double-blind reviewing, will only be integrated once the paper has been accepted for publication.}

To summarize, this paper presents the following  contributions:
\begin{itemize}
	\item the Android App Analysis Query Language (\AQL), a mechanism to precisely define taint-analysis queries and responses,
	\item the \emph{\AQL-System}, which dispatches \AQL queries to tools and consolidates their responses,
	\item the \emph{Benchmark Refinement and Execution Wizard} (\BREW), a tool to refine, execute and precisely evaluate formulated benchmarks, and
	\item a large-scale, comparative empirical evaluation of \Amandroid, \DIALDroid, \DidFail, \DroidSafe, \FlowDroid and \IccTA on \DIALDroidBench, \DroidBench, \ICCBench and 21 newly developed test apps.
\end{itemize}

%Furthermore, we present the outcome determined by \BREW~for the most-commonly used benchmarks and well-known taint analysis tools.
%Along with that, we explain in detail why benchmark improvements are needed and contribute such an improved benchmark, namely \emph{\AQL-Bench}.\\

The remainder of this paper is structured as follows.
Section~\ref{sec:background} introduces some basic concepts and a running example.
Section~\ref{sec:approach} details \tool's three major components.
Section~\ref{sec:evaluation} and \ref{sec:results} present our large-scale comparative evaluation and its results for the six mentioned taint analysis tools.
We discuss related work in Section~\ref{sec:related_work} and conclude in Section~\ref{sec:conclusion}.

%% file: content/background.tex
In this section, we start with introducing basic terminology and concepts, 
in particular explain taint analysis and the current form of benchmark suites. 

%In this section we first present prominent, mostly Android specific security threats.
%Second, we briefly describe a program analysis type that represents 
%a countermeasure to these threats.
%Third, we introduce benchmarks, which are used to measure the effectiveness of analyses.
%Along the way, we highlight issues of currently existent analyses and benchmarks that can be overcome by the approach purposed in Section~\ref{sec:approach}.
%Furthermore, a running example is started here and continued in the following sections.

\subsection{Taint Analysis}\label{subsec:taintanalysis}
The purpose of taint analysis is to track the flow of sensitive data within programs. 
For smart-phone apps, a data leak occurs when private data (phone numbers, device identifiers, contact data) flows from sensitive sources to public sinks (Internet, SMS transmission). 
In this case, sensitive data is leaked. % -- unless the app has an explicit {\em permission} to do so. 
%A frequently used technique to detect data leakage is taint analysis: 
Taint analyses are most frequently used to detect such leaks:
it taints sensitive data at its source, and propagates the taint information through the application (or even a combination of apps), issuing a warning if tainted data reaches a sink.
Taint tracking can be performed statically on program code, or dynamically by executing the app and monitoring tainted data. 
%For the purpose of this reproducibility study we focus on static taint analysis.

Android provides an open communication structure between apps.
Moreover, when Android apps include third-party libraries, those execute with the same access rights as the app itself. 
Those features make Android apps particularly vulnerable to attacks targeting private data. 
Taint-analysis tools can cope with these special features to various extents. 
The following programming-language features and analysis functionalities are supported by some but not all tools:
%\begin{description}
%	\item[Context-Sensitivity]
%	\item[Field-Sensitivity]
%	\item[Object-Sensitivity]
%	\item[Reflection] Java's reflection mechanism allows to (statically) 
%             hide the names of called methods. 
%	\item[ICC and IAC] 
%             Android allows for inter-component or inter-app communication (ICC/IAC) via 
%             the instantiation of so-called \emph{Intents} and \emph{Intent-Filters}.  
%             For example, to access the device's camera it is enough to instantiate a certain Intent. 
%	\item[...]
%\end{description}
\begin{description}
	\item[Aliasing] {
		The same memory location / object may be referenced by different variables.
		In this situation, one variable is an alias of another, and a taint related to one alias must be carried over to all others.
		}
	\item[Static Fields] {
		Static fields are declared on a type, not their instance.
		In particular, their values can be accessed without requiring access to any object reference.
		Static taint analyses must thus treat static fields differently from instance fields.
	}
	\item[Lifecycle and Callbacks] {
		Each Android component has its own lifecycle, defining a sequence of invocations to callback functions that the Android framework issues at appropriate lifecycle events.
		User-interface interactions map to callbacks as well.
		To model all possible execution sequences of an app, the analysis must take all appropriate callbacks into account, and it can do so with various levels of precision.
	}
	\item[Inter-Component Communication] {
		A leak may originate in one class and end in another.
		Additionally, Android allows for inter-component or inter-app communication (ICC/IAC) via the instantiation of so-called \emph{intents} and \emph{intent-filters}.  
		For example, to access the device's pre-installed camera app, it is sufficient to dispatch a certain intent.
		Intents may propagate tainted data from one app or component to another.
	}
	\item[Analysis Abstraction and Algorithmics] {
		Depending on the exact analysis abstraction and algorithmics, the taint analysis may or may not support different ``sensitivities'', such as flow, context, path, field, object and/or thread-sensitivity~\cite{DBLP:conf/pldi/ArztRFBBKTOM14}.
		While generally, the support for more such ``sensitivities'' may increase precision, reducing the amount of false positives, the positive effects differ.
		For instance, while some level of object-sensitivity is known to be important for the precise analysis of Java and Android applications~\cite{DBLP:conf/popl/SmaragdakisBL11}, thread-sensitivity may well be less important in the case of Android. 
	}
	\item[Reflection] {
		Java's reflection mechanism allows one to invoke methods (or access fields) through dynamically generated strings.
		An analysis must resolve these strings to reliably detect taint flows through such invocations.
	}
\end{description}

The fact that each Android taint analysis tool supports those features to a different extent makes it important to evaluate them comparatively, as without such comparison it is impossible to tell which features actually matter.
Conducting a fair and automatic comparison of tools, however, is complicated by differences in pursued target flows and output formats. 
A necessary part of every taint analysis is the identification of sources, i.e., statements that extract sensitive information from a certain resource, and sinks, i.e., statements that exfiltrate information out of the app.
There exist different definitions of sources and sinks as well as various approaches to determine which statements belong to which set~\cite{DBLP:conf/ndss/RasthoferAB14, DBLP:conf/ccs/AuZHL12}.
For instance, some analysis tools consider logging statements as sinks, others do not.
Whereas one analysis approach defines that each source and sink has to be protected by at least one permission, another might not consider such constraints. 

The second difference concerns the representation of flows between sources and sinks, for which every tool uses its own (usually textual, individually-structured) format.
This renders it impossible to automatically compare results produced by different tools without additional effort.
For example, the result generated by \textsc{FlowDroid} (see Listing~\ref{lst:flowdroid}), 
from a structural point of view, has little in common with \textsc{DidFail}'s result 
(see Listing~\ref{lst:didfail}).
Yet, both listings show the same finding:
a flow from \texttt{getDeviceId()} (source) to \texttt{sendTextMessage(...)} (sink).

\begin{lstlisting}[float, keywords={getDeviceId, sendTextMessage}, 
                  label=lst:flowdroid, caption=Excerpt of \textsc{FlowDroid}'s result (\texttt{DirectLeak1.apk})]
Found a flow to sink virtualinvoke $r4.<android.telephony.SmsManager: void sendTextMessage(java.lang.String,java.lang.String,java.lang.String,android.app.PendingIntent,android.app.PendingIntent)>("+49 1234", null, $r5, null, null), from the following sources:
        - $r5 = virtualinvoke $r3.<android.telephony.TelephonyManager: java.lang.String getDeviceId()>() (in <de.ecspride.MainActivity: void onCreate(android.os.Bundle)>)
\end{lstlisting}

\begin{lstlisting}[float, keywords={getDeviceId, sendTextMessage}, 
                  label=lst:didfail, caption=Excerpt of \textsc{DidFail}'s result (\texttt{DirectLeak1.apk})]
### 'Sink: <android.telephony.SmsManager: void sendTextMessage(java.lang.String,java.lang.String,java.lang.String,android.app.PendingIntent,android.app.PendingIntent)>': ###
['Src: <android.telephony.TelephonyManager: java.lang.String getDeviceId()>']
\end{lstlisting}

\subsection{Benchmarks}\label{subsec:benchmarks}
Most Android app analysis tools are evaluated by means of benchmarks.
In this context, a benchmarks is a collection of apps that have certain features.
For example, the most frequently used micro-benchmark 
\textsc{DroidBench}~\cite{DBLP:conf/pldi/ArztRFBBKTOM14} comprises 190 apps.
%Each of these apps exploits a certain feature of Java or Android.
%In doing so, numerous privacy leaks are constructed.
%To make use of \textsc{DroidBench}, we first have to choose an analysis tool.
%Then we execute it in order to analyze all these 190 apps. \todo{Heike}{warum muss das hier erklaert werden?}\todo{Felix}{Weil ein benchmark kein program ist, das ich nur starten brauch.}
%Last, we can rate the tool depending on how many \textit{correct} privacy leaks were found.
%
\begin{lstlisting}[language=java, float, label=lst:droidbench, caption=Excerpt of DroidBench's app source code (\texttt{DirectLeak1.apk})]
import android.telephony.TelephonyManager;
/**
 * @testcase_name DirectLeak1
 * @version 0.1
 ...
 * @description Easy testcase: The value of a source is directly sent to a sink
 * @dataflow source -> sink
 * @number_of_leaks 1
 * @challenges -
 */
public class MainActivity extends Activity {
\end{lstlisting}
While most Android taint analysis tools are evaluated by applying them to \textsc{DroidBench}, such evaluations are of limited value because \DroidBench currently only imprecisely specifies the {\em ground truth} in each benchmark case, i.e., the correct result of flow analyses.
Listing~\ref{lst:droidbench} gives an excerpt of the source code of one of \textsc{DroidBench}'s apps, namely \emph{DirectLeak1} (the one the prior analysis results referred to). 
Between the imports and the source code of the main activity of this app a comment is placed. 
This comment -- including the number of leaks -- is the only description of the expected analysis results: there is a source directly connected to a sink. 
Neither do we learn which source and sink this might be, nor how the flow propagates through the app(s). 
Moreover, the information is not given in a machine-readable format. 
%It holds information about the case's name, version and author as well as 4 elements that describe what should be found in this app.
%The \texttt{@description} stored in this comment tells us that the source is directly connected to the sink.
%One line later this information is provided again in a shorter way (\texttt{@dataflow}).
%Additionally, challenges which have to be overcome in order to successfully detect all taint flows in this app are mentioned.
%In this particular case there are no specific challenges (\texttt{@challenges -}).
%The only machine-readable information can be found next to the \texttt{@number\_of\_leaks} annotation.
%In this particular case only 1 information flow should be findable.
%More precise information on this flow is missing.
Looking it up in the source code reveals that the source is a \texttt{getDeviceId()} function call, 
which is used as a parameter of a \texttt{sendTextMessage(...)} statement.
In this case, a manual inspection is easy and can be done quickly. 
However, for more challenging benchmark cases or hundreds of cases this task becomes exhausting and error prone.
For example, the machine readable information, number of leaks, is sometimes simply wrong (e.g.\ for \DroidBench's \emph{StrongUpdate1} it says 1 but the description and app execution prove that there is no leak).
%Specialized tools as described in Section~\ref{sec:approach} are needed to reduce the effort.
Still, the authors of \textsc{ICC-Bench}~\cite{DBLP:conf/ccs/WeiROR14} adopted the same method to 
specify their expected results.
Hence, automatically comparing actual results against expected ones is neither possible for 
\textsc{DroidBench} nor for \textsc{ICC-Bench}.

\textsc{DroidBench} and \textsc{ICC-Bench} are micro-benchmarks. 
Evaluations on real-world apps usually lack a ground truth altogether.
Such evaluations usually focus on the 500 to 1000 most downloaded or most popular apps that can be downloaded from Google's PlayStore.
\textsc{DIALDroid-Bench}~\cite{DBLP:conf/ccs/BosuLYW17} is a benchmark suite comprising real-world apps. 
Its apps, however, are given without source code, only in the form of .apk-files, and comprise no description of the data leaks they contain.

In summary, while there seems to be a general agreement of evaluating tools on the grounds of public benchmarks, the evaluation itself is imprecise:
The ground truth needed to determine the tools' precision and recall is often simply not known. 
To evaluate tools automatically, one further misses a standardized, machine-readable format for expected as well as detected data flows. 
%
%In summary, there exists a common method to determine and evaluate experimental benchmark results.
%However, we cannot do this automatically since a clear, machine-readable description of the ground truth is unavailable.
%Additionally, there exists no standardized format for the output of analysis tools.
%When it comes to real world apps we often do not even know the set of apps that has been used in certain experiments.
%In the end, this disables everyone "to reproduce the analysis or even replicate the study in the same or a similar context."~\cite{DBLP:journals/tse/KitchenhamPPJHER02}
In the end, this impedes a reproducible and unbiased comparison of analysis tools. 
We next explain how \tool overcomes these limitations.

%% file: content/approach.tex
The \tool framework supports tool evaluation and comparison by the following three concepts: 
\begin{itemize}
	\item the design of \AQL, a \emph{query language} for precisely formulating  questions about app properties such as flows, 
	\item the design and implementation of a \emph{query execution system} being able to interface 
		to diverse tools, and 
	\item  the design and implementation of a \emph{query wizard} for determining the ground truth in apps and for executing thus specified benchmarks and rating their outcome. 	
\end{itemize}
\noindent Together, they provide an automatic benchmarking system for app-analysis tools.
In the following, we describe all three parts in turn. 

%The approach described in this chapter can be used to tackle the previously described issues.
%It represents one part of our contribution and consists of three parts:
%First, the \emph{Android App Analysis Query Language (\AQL)}.
%Second, an associated system called \emph{\AQL-System}.
%And last, \emph{\BREW}, the \emph{Benchmark Refinement and Execution Wizard}.
%Together these three parts allow us to execute arbitrary benchmarks with arbitrary analysis tools.
%All results computed this way can be evaluated automatically, with regards to precision, recall and f-measure.
%Furthermore, the results can automatically be compared with each other.
%In the following, each of the three parts are explained in more detail.
%At the end of this chapter an overview over the whole approach is provided.

\subsection{App Analysis Language}\label{subsec:AQL}
The Android App Analysis Query Language (\AQL)~\cite{Pauck_2017} consists of two main parts, 
namely \AQL-Queries and \AQL-Answers.
\AQL-Queries enable us to ask for Android specific analysis subjects in a general, tool independent way.
The grammar defining \AQL-Queries currently allows to ask for analysis subjects such as flows, intents, intent-filters and permissions.
Considering our running example, we can enumerate all taint flows within the \texttt{DirectLeak1.apk} app by composing the following query:
\begin{center}
	\texttt{\textbf{Flows IN} App('/path/to/DirectLeak1.apk') \textbf{?}}
\end{center}
or instead explicitly check the taint flows we expect:
\begin{center}
	\texttt{\textbf{Flows FROM}\\Statement('getDeviceId()')\\->Method('onCreate(...)')->Class('MainActivity')\\->App('/path/to/DirectLeak1.apk')\\\textbf{TO}\\Statement('sendTextMessage(...)')\\->Method('onCreate(...)')->Class('MainActivity')\\->App('/path/to/DirectLeak1.apk')\\\textbf{?}}
	% \texttt{\textbf{Flows FROM}\\Statement('getDeviceId()')\\->Method('onCreate(...)')\\->Class('MainActivity')\\->App('/path/to/DirectLeak1.apk')\\\textbf{TO}\\Statement('sendTextMessage(...)')\\->Method('onCreate(...)')\\->Class('MainActivity')\\->App('/path/to/DirectLeak1.apk')\\\textbf{?}}
\end{center}

%Thereby, analysis tools capable of answering such detailed questions can work more efficiently.
Furthermore, the \AQL~ offers several options to merge and filter queries as well as methods to match intents and intent-filters.
%However, in the context of our approach described here, we only treat it as a mean to ask arbitrary analysis tools for their results.
Similarly, \AQL-Answers are used to represent analysis results in a standardized form. 
The syntax of \AQL-Answers is defined via an XML schema definition. 
%Considering such results the \AQL-Answers come into play.
%\AQL-Answers are used to represent analysis results generalized but without loss of precision.
%Thus, all elements we can ask for with \AQL-Queries can be represented by means of \AQL-Answers.
%The structure is syntactically defined by an XML schema definition.
%Thereby, \AQL-Answers can be constructed easily, parsed with automatically-generated parsers and validated against that schema.
Considering the running example again, we might get an answer that represents the 
flow from \texttt{getDeviceId()} to the \texttt{sendTextMessage(...)} statement (cf. Listing~\ref{lst:aqlanswer}).
In this, each statement comes with a precise description of where it can be found, by naming the method, class and app containing the statement. 
%To this end, the method, class and app that holds this statement is named.
\AQL supports additional syntax to uniquely identify statements and apps, for example using function-call parameters, full Jimple syntax\footnote{Jimple stands for "Java but simple". It is the primary intermediate language supported by Soot~\cite{DBLP:conf/cascon/Vallee-RaiCGHLS99, cetus11soot}.} statements or \verb+.apk+ file hashes.

\begin{lstlisting}[language=xml, keywords={answer, flows, flow, reference, statement, method, classname, app, file, hashes}, float, 
                  label=lst:aqlanswer, caption=Shortened \AQL-Answer (\texttt{DirectLeak1.apk})]
<answer>
    <flows>
        <flow>
            <reference type="from">
                <statement>... getDeviceId() ...</statement>
                <method>... onCreate(...) ...</method>
                <classname>... MainActivity</classname>
                <app>
                    <file>.../DirectLeak1.apk</file>
                    <hashes>...</hashes>
                </app>
            </reference>
            <reference type="to">
                ...
                sendTextMessage(...)
                ...
            </reference>
        </flow>
    </flows>
</answer>
\end{lstlisting}

\subsection{Query Execution System}\label{subsec:AQLSystem}
The \AQL-System~\cite{TL:AQL-System} is our approach to process \AQL-Queries and to determine \AQL-Answers.
In the scope of this paper it is sufficient to observe the \AQL-System as a blackbox, which accepts analysis questions encoded in \AQL-Queries as input, executes appropriate analysis tools and converts their output into \AQL-Answers.
%This blackbox must be configured via a configuration file in XML format giving information about
To do so, it requires a \emph{configuration} in form of an \verb+.xml+ file that describes 
(a) which tools are avaliable in a certain instance of the \AQL-System and how to execute these, (b) which queries can be answered by which tool and (c) how to convert a tool's result into an \AQL-Answer.
For instance, an AQL-System can be configured to execute \FlowDroid in case of intra-app flow questions and \IccTA in case of inter-app questions, since \FlowDroid does not support ICC/IAC.
Considering the running example such an \AQL-System recognizes that \FlowDroid is available and able to answer the query regarding flows inside one app only.
Consequently, \FlowDroid is launched by executing the command or script specified in the configuration.
Once its computation is finished a tool-specific converter translates the tool's result into an \AQL-Answer.
%The \AQL-System provides an interface to arbitrary tools by translating requested \AQL-Queries into \emph{configurations} \fbox{ja?} for analysis tools and their results back into \AQL-Answers. 
%It does so by means of a configuration file in XML format that describes 
%which tools are available in a certain instance of the \AQL-System and how to convert queries and answers. 
We currently have converters covering in particular the six tools of our experiments in Section~\ref{sec:evaluation}. 
With this, we can orchestrate the execution of tools and convert their results into the standardized \AQL~ format.
%With this, we can automatically steer execution of the tools and convert their results back into the standardized format. 
%With the help of this knowledge the \AQL-System is able to produce an \AQL-Answer once it receives an \AQL-Query by executing one or more analysis tools.
%Continuing the running example, we could write a configuration file that specifies, among other tools, a taint analysis tool such as \FlowDroid.
%This file then holds information about how to execute \FlowDroid and which questions it may answer, in case of \FlowDroid intra-component Flow questions.
%In addition, more details can be given, for example, how much memory is provided to \FlowDroid.
%Once the \AQL-System configured this way receives a query, which asks for Flows, it executes \FlowDroid and converts its result into an \AQL-Answer at the moment it becomes available.
%To do so, it requires a tool specific converter.

%The \AQL~ as well as the \AQL-System also offers a wide range of additional features from requesting preprocessors in a query over setting up a priority queue for tools in case the execution of one tool may fail down to specifying maximal execution times for tools.

\subsection{Benchmark Refinement and Execution}\label{subsec:BREW}
\tool's final component is the Benchmark Refinement and Execution Wizard (\BREW)~\cite{TL:BREW}.
It is an assistant that can be used to do what the name suggests, first refine and then execute a benchmark.
For this, it offers a graphical user interface (GUI) simplifying the handling of different sets of apps and benchmarks 
and the identification of sources and sinks. 
%As well as a command line interface to be usable on any system.

The process of refining an existing benchmarks suite, i.e., completing it and bringing it into standardized format, 
consists of three steps: 
\begin{description}
	\item[Case Identification] We load the apps of the test suite into the wizard. 
	After loading, each app resembles one benchmark case.
	We can then restructure the cases by deactivating certain cases or combining cases (e.g., if we are interested in a flow from a source in one app to a sink in another app).
	\item[Source and Sink Identification] For source and sink identification, 
             the wizard displays all statements that possibly represent a source or a sink inside all apps 
	      that have been loaded during the first step.
		Since the wizard is just an assistant it cannot decide on its own which of these statements are real sources and sinks.
		However, \BREW~can preselect all sources and sinks according to a predefined list of sources and sinks as produced by SuSi~\cite{DBLP:conf/ndss/RasthoferAB14} or PScout~\cite{DBLP:conf/ccs/AuZHL12}.
		It is up to the user to perform the remaining task of deselecting all unwanted sources and sinks from the preselected ones. 
		It is also possible to combine certain sources and sinks. This might be necessary to unify differently defined sources and sinks. 
For example, whereas one analysis considers the function call \texttt{getLastKnownLocation(...)}, which returns an \texttt{Location} object, as source, another analysis only considers the call of \texttt{getLongitude()} or \texttt{getLatitude()}, called upon such a \texttt{Location} object, as source.
		However, any of these calls refers to the same resources and hence all calls can be interpreted as a single source.
	\item[Ground Truth Identification] Finally, we have to decide which cases are \emph{positive} and 
			which cases are \emph{negative} benchmark cases.
		More precisely, there may exist cases where we define a flow that should not be found by an analysis.
             Determining positive and negative cases remains a manual task which requires inspection of the case. 
		%Hence, such a case should be marked as a negative benchmark case representing a candidate for a false positive answer.
\end{description} 

Considering the running example only the \texttt{DirectLeak1.apk} may be loaded as first step.
Then, if we choose to automatically mark all sources and sinks according to \SuSi, the \texttt{getDeviceId()}, \texttt{sendTextMessage(...)} but no further statement get marked as source and sink.
In the last step this results in one benchmark case, which is correctly and initially always marked as positive case by \BREW.

Once the refinement steps have been completed, the benchmark can be executed and evaluated.
To do so, \BREW~ determines one \AQL-Query and one (expected) \AQL-Answer per benchmark case.
%On the one hand, the answer represents the expected outcome of an analysis with respect to its associated benchmark case.
%On the other hand, the query can be used together with an \AQL-System to execute an analysis tool in order to compute an actual analysis result also in form of an \AQL-Answer.
Hence, to automatically execute and evaluate a benchmark, \BREW~ sends a query to an \AQL-System for each benchmark case 
and checks whether the actual result determined that way matches the expected one.
For this purpose, it is checked whether one flow of the expected result matches one flow of the actual result.
For example, one analysis may detect a flow from \texttt{getLastKnownLocation(...)} to \texttt{sendTextMessage(...)} 
whereas another analysis finds a flow from \texttt{getLongitude()} to the same sink.
In both cases the expected flow, regarding the accessed resources, has been found.
Consequently, one matching flow per benchmark case is sufficient.

To evaluate the outcome of a benchmark execution, 
\BREW~ counts the number of successful and failed benchmark cases.
A case is \emph{successful} if a certain flow that was expected to be found has been found (true positive) 
or if a flow that was explicitly not expected to be found has not been found (true negative).
In contrast, a case \emph{fails} if an expected flow was not found (false negative) or if a not expected flow has falsely been detected (false positive).
Based on this information, \BREW computes the commonly used metrics precision, recall and F-measure.

Overall, \BREW~ helps and guides its users while refining a benchmark.
Refining in this context refers to the process of adding missing information to a benchmark case.
Thereby insufficiently described benchmark cases (cf. Listing~\ref{lst:droidbench}) become usable, i.e., become available in machine-readable format, which allows one to execute and evaluate the benchmark cases automatically.

%The two uses of \BREW, refining and executing, can be done in a distinct manner as well.
%For example, the authors of a benchmark can perform the refinement step.
%Whereas anyone else can use such a refined benchmark directly to, for instance, evaluate their tool.

\subsection{The \tool Toolchain}\label{subsec:toolchain}

\begin{figure}
	\centering
	\includegraphics[width=\columnwidth]{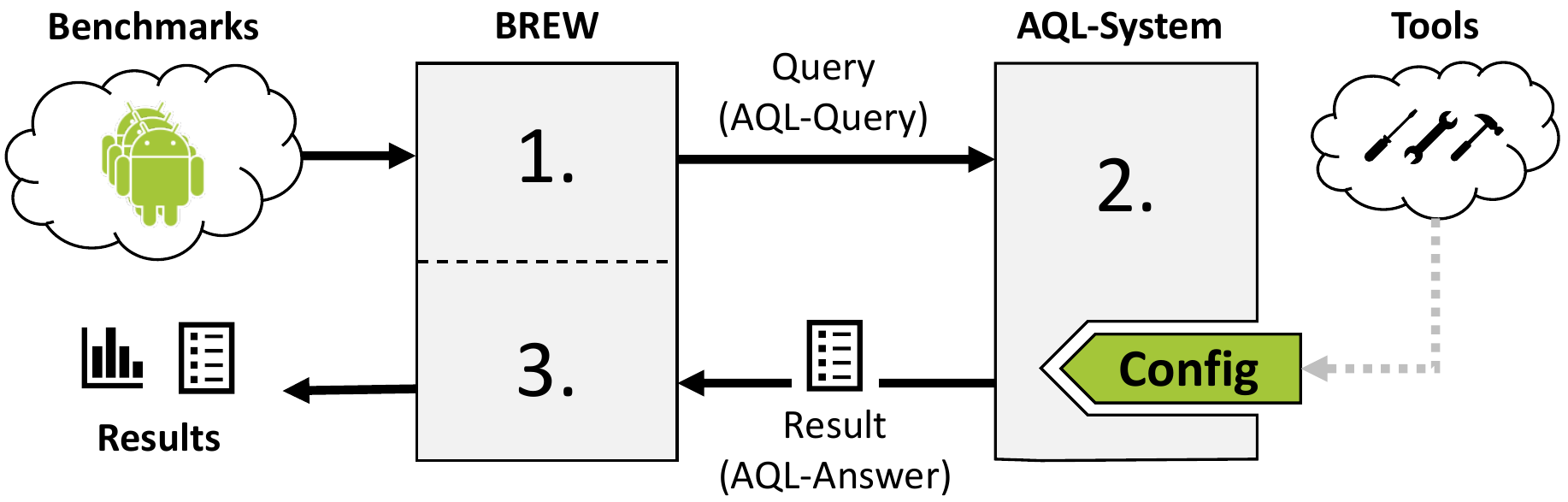}
	\caption{Sketch of the \tool toolchain}
	\label{fig:overview}
\end{figure}

The composition of the three previously described components form our complete framework \tool (see Figure~\ref{fig:overview}):
\BREW~ uses the \AQL-System which again uses the \AQL.
In Figure~\ref{fig:overview}, two parts are visualized inside clouds. 
These symbolize the resources that exist in the community, namely analysis tools and benchmarks.
Each gray rectangle represents one of \tool's components.
To execute and evaluate a benchmark we first have to choose one and provide it to \BREW~as input.
Two methods are available to do so.
Either an already refined benchmark can be loaded, or a new, unrefined set of apps can be refined interactively.
%This is the first task \BREW~ deals with as visualized by \emph{1.} in Figure~\ref{fig:overview}.
After setting up or loading the desired benchmark, \BREW~ will issue one \AQL-Query per benchmark case.
Then, one query after another arrives at an \AQL-System, configured to use a certain set of tools.
The \AQL-System executes the tools required to answer each query (\emph{2.}) and produces one \AQL-Answer per query as output.
This answer, in turn, is returned to \BREW, which then decides if this actual answer matches the expected one (\emph{3.}).
\BREW's GUI provides possibilities to compare single results on a textual or graphical level.
The latter one depicts all sources and sinks as nodes and connections between them as edges in a graph.
All known intermediate statements between each source and sink are also depicted.
\BREW further allows one to export complete results to an SQL database, e.g.\ to make them viewable online.
In summary, we input a benchmark and one or more analysis tools and receive as output one benchmark result, which includes the expected and actual \AQL-Answers of each benchmark case, as well as the calculated values for precision, recall and F-measure.

Considering the running example we could, for instance, input the \texttt{DirectLeak1.apk} 
as a benchmark consisting of only one app and setup an \AQL-System to use \FlowDroid.
The refinement step performed by means of \BREW~ allows us to identify the expected leak 
which is consequently available as ground truth or expected \AQL-Answer (see Listing~\ref{lst:aqlanswer}).
Along with that, an \AQL-Query is composed to ask whether the expected leak actually exists.
The considered \AQL-System takes this query and thereupon executes \FlowDroid.
Once \FlowDroid finishes its computation, the \AQL-System converts \FlowDroid's result 
(see Listing~\ref{lst:flowdroid}) into an \AQL-Answer (see Listing~\ref{lst:aqlanswer}) which is returned to \BREW.
Since the expected and actual result coincide (also see Listing~\ref{lst:aqlanswer}), 
the benchmark case is evaluated as successful.
Considering this tiny benchmark, precision, recall and F-measure would be at its optimal value ($1.0$).

%This toolchain can be used to overcome the two issues described before:
%Multiple sources and sinks can be merged, thus different source-sink definitions can be handled (see Section~\ref{subsec:tools}).
%By refining benchmarks the ground truth becomes available and is not unknown or insufficiently described anymore (see Section~\ref{subsec:benchmarks}).
%A method to overcome these issues was needed for a long time, because many studies executed benchmarks such as DroidBench in an inappropiate manner.
%With this toolchain this should not be possible to happen again.
%Furthermore, this toolchain largely simplifies the execution of benchmarks.
%In the next section, this toolchain is used to reevaluate certain tools in order to answer questions such as: Do Android app analysis tools keep their promises?

This toolchain is applicable to different scenarios.
For instance, a specialist could refine a benchmark based on his expertise.
This refined benchmark can then be executed and used in an evaluation by any type of user such as an analysis tool developer.
The toolchain also allows us to refine and execute a benchmark in a distributed and iterative way.
Multiple people can iteratively refine a single benchmark.
A benchmark can be executed, extended by some additional cases and executed again without rerunning the already executed cases.
This can split the burden of refining and executing a benchmark among multiple persons.\footnote{Instructions are available on github: \url{https://github.com/FoelliX/BREW/wiki}}

%% file: content/evaluation.tex
The design and implementation of \tool for the first time facilitates an accurate comparison of tools. 
In the following, we re-evaluate six taint analysis tools for Android apps and their promises on a number of micro-benchmarks and real world apps. 
%In this section we deal with 3 research questions regarding the defensibility of promises made by 6 Android app analysis tools (see Table~\ref{tab:tools}).
%To do so, we present different experimental setups handling
%{\color{white}\footnote{\label{fn:amandroid}\url{https://bintray.com/arguslab/maven/argus-saf/3.1.2}}\footnote{\label{fn:dialdroid}\url{https://github.com/dialdroid-android/DIALDroid}}\footnote{\label{fn:didfail}\url{https://www.cert.org/secure-coding/tools/didfail.cfm}}\footnote{\label{fn:droidsafe}\url{https://mit-pac.github.io/droidsafe-src/}}\footnote{\label{fn:flowdroid}\url{https://github.com/secure-software-engineering/soot-infoflow-android/wiki} (Commit-ID: 5696905)}\footnote{\label{fn:iccta}\url{https://sites.google.com/site/icctawebpage/source-and-usage}}}
% Bsp: $^\text{\ref{fn:amandroid}}$

\subsection{Tool Selection}
We start with a description of our tool selection.
The tools selected for our evaluation all implement {\em taint} analyses. 
This is the most frequently used technique for finding data leaks,
 and thus provides us with a number of tools to be studied. 
%since we are mainly interested in privacy leaks, these are the most promising.
Furthermore, for the purpose of this study we only consider {\em static} taint analysis tools. 
We further consider only approaches that are at least flow-sensitive or context-sensitive, such as to assure that all tools are competitors within the same league.
%On the one hand, it is often undescribed which apps belong to one benchmark case.
%On the other hand, the results are even more imprecise since we regularly do not get to know to which app a flow, source or sink belongs.

\begin{wraptable}[10]{l}{4.2cm}
%\begin{table}[t]
	%\caption{Tools considered in the evaluation}
	\caption{List of Tools}
	\label{tab:tools}
	\scalebox{0.8}{
		\begin{tabular}{|c|c|}
			\hline
			\textbf{Tool} & \textbf{Version} \\
			\hline
%			\Amandroid~$^\text{\ref{fn:amandroid}}$	& November 2017 (3.1.2) \\
%			\DIALDroid~$^\text{\ref{fn:dialdroid}}$	& September 2017 \\
%			\DidFail~$^\text{\ref{fn:didfail}}$		& March 2015 \\
%			\DroidSafe~$^\text{\ref{fn:droidsafe}}$	& June 2016 (Final) \\
%			\FlowDroid~$^\text{\ref{fn:flowdroid}}$	& April 2017* (Nightly) \\
%			\IccTA~$^\text{\ref{fn:iccta}}$			& February 2016 \\
			\Amandroid~{\small \cite{TL:Amandroid}}	& November 2017 (3.1.2) \\
			\DIALDroid~{\small \cite{TL:DIALDroid}}	& September 2017 \\
			\DidFail~{\small \cite{TL:DidFail}}		& March 2015 \\
			\DroidSafe~{\small \cite{TL:DroidSafe}}	& June 2016 (Final) \\
			\FlowDroid~{\small \cite{TL:FlowDroid}}	& April 2017* (Nightly) \\
			\IccTA~{\small \cite{TL:IccTA}}			& February 2016 \\
			\hline
			\multicolumn{2}{c}{{\small* Has been updated since then.}}
		\end{tabular}
	}
\end{wraptable}
Table~\ref{tab:tools} lists the tools employed in our evaluation along with their release dates and versions (if available). 
For all tools we generally used the most recent version. 
Just for \FlowDroid it holds that it is so frequently updated that we had to fix a version for our experiments.
All six tools we consider have at least ICC and at best IAC capabilities, except for \FlowDroid, which computes flows within single components only.
Yet, it is an important tool to consider because it is the most widely used static-analysis tool for Android.
Considering IAC is interesting because prior evaluations typically showed a deteriorating precision when IAC benchmark cases were involved.
We thus seek to specifically measure how well tools perform on more accurate IAC benchmarks.

We briefly describe some characteristics of the tools:
\begin{description}
 \item[Analysis Engine]
All tools except \Amandroid are based on Soot~\cite{DBLP:conf/cascon/Vallee-RaiCGHLS99, cetus11soot} 
and operate on Jimple as intermediate language.
\item[Source and Sink Identification] 
%Of further importance for a comparison is the technique used to identify {\em sources} and {\em sinks}. 
The sources and sinks considered by \DIALDroid, \DidFail, \FlowDroid and \IccTA are specified by \SuSi~\cite{DBLP:conf/ndss/RasthoferAB14}, 
a machine-learning approach for source and sink detection. 
\DroidSafe employs its own source and sink identification (which is claimed to be even more precise than \SuSi's). 
The list of considered sources and sinks used by \Amandroid seems similar although shorter; its origin remains unclear. 
For our micro-benchmarks, we made sure that the sources and sinks needed for finding flows are identified by all tools. 
\item[ICC and IAC Capabilities] \DidFail, \DIALDroid and \DroidSafe are the only tools that are shipped with built-in IAC capabilities.
The other tools have ICC capabilities only and require a tool called \ApkCombiner~\cite{DBLP:conf/icse/0029BBKTARBOM15} to lift their analysis to IAC level.
\ApkCombiner has been developed (along with \IccTA) for precisely this purpose.
It takes multiple \verb+.apk+ files as input and merges them into a single \verb+.apk+ file.
\DroidSafe's built-in IAC capabilities did not show any effect in our experiments, no IAC involving flows were found.
Hence, we decided to use \DroidSafe in combination with \ApkCombiner as well.
Note furthermore that \DIALDroid is only able to detect inter-component or inter-app taint flows, any intra-component flows are ignored.
\end{description}

\noindent A number of other tools (e.g.~\cite{DBLP:conf/sigsoft/FengDAA14, DBLP:conf/wisec/CuiWHXZY15, DBLP:conf/dsn/BagheriSBM16, scandroid, DBLP:conf/issta/HuangDMD15}) would fit into our scope.
We shortly comment on and provide reasons why we omitted them in related work (see Section~\ref{sec:related_work}).
%Most of them, however, are not publicly available and have been omitted for this reason.
%In future we will attempt to get more tools and include them into our approach and evaluation.

\subsection{Benchmarks}
Our experiments are based on three benchmark suites: 
\DroidBench~\cite{DBLP:conf/pldi/ArztRFBBKTOM14}, \ICCBench~\cite{DBLP:conf/ccs/WeiROR14} and \DIALDroidBench~\cite{DBLP:conf/ccs/BosuLYW17}.
The first two are well-known micro-benchmark suites which have been used in various evaluations before 
(usually version 2.0 or 3.0 of \DroidBench and version 2.0 of \ICCBench which we use as well). 
%Hence, we also consider these versions.
The third suite, \DIALDroidBench, is a collection of partially malicious real-world apps downloaded from Google's Playstore and 
gathered by Bosu et al.~\cite{DBLP:conf/ccs/BosuLYW17}.

In addition, we have developed 18 apps comprising 21 positive and 6 negative \emph{feature-checking} benchmark cases.
A feature-checking benchmark case exploits only one specific feature at a time and 
can thus be used to explicitly check the handling of a dedicated feature in a tool. 
This is in contrast to similar cases of \DroidBench which often combines two or more features in one case. 
%, these exploit only one feature at a time and thus can be used to check whether individual functionalities are supported.
The feature-checking benchmark cases cover all features listed in Section~\ref{subsec:taintanalysis}.

Since we are particularly interested in ICC and IAC, we have developed three apps to specifically evaluate the precision of \emph{intent-matching} algorithms.
Such algorithms play an essential role when inter-component or inter-app flows are analyzed.
The analysis has to detect whether a certain intent can be received by a component.
If so, the action, category and data attributes of an intent have to match those of a component's intent-filter.
To this end, the three apps comprise 2/6/71 positive and 4/3/139 negative cases 
considering matching action, category and data attributes, respectively.  

Together these 21 newly developed apps represent our \DroidBench extension.
\DroidBench together with this extension is refined by means of \BREW.
As a result, our benchmark suite now contains a collection of 211 apps with (a) their source code and 
(b) the ground truth for data leaks in the form of \AQL-Answers. 
%, the app and its source code is available.
We made this extended and refined benchmark suite as well as refined versions of the other suites publicly available~\cite{TL:ReproDroid} for other researchers to perform similar experiments.

\subsection{Research Questions and Experiments}
\newcommand{\RQONE}{Do Android app analysis tools keep their promises?}
\newcommand{\RQTWO}{How do the tools compare to each other with respect to accuracy?}
\newcommand{\RQTHREE}{Which tools support large-scale analyses of real-world apps?}
%state research questions and evaluation plan
Our evaluation addresses the following research questions:
%Various experiments are executed to answer the following research questions.
%Whereas their experimental setup is described next their results are discussed in Section~\ref{sec:results}.
\begin{enumerate}
	\item[\textbf{RQ1}] \RQONE
	\item[\textbf{RQ2}] \RQTWO
	%\eb{Das können wir so nicht schreiben, denn wir messen das ja nur einige wenige Faktoren.}
	\item[\textbf{RQ3}] \RQTHREE
\end{enumerate}

\noindent We designed specific experiments for every research question. 

\textbf{RQ1}. In order to address RQ1, we first need to determine what we consider to be a ``promise'' of a tool. 
Looking at the articles introducing tools, two properties of tools play a common role: 
(1) the supported features and (2) the tool's accuracy, i.e., its precision, recall and F-measure as shown in experiments. 
Runtime appears to play a minor role: most articles  only give vague runtime information.
To evaluate the tools with respect to these sorts of promises, we ran the following experiments. 
First we prepared a benchmark set consisting of
(1)  the micro-benchmarks from \DroidBench and \ICCBench plus 
(2) our \emph{feature-checking} benchmark cases, all refined and executed with \BREW.
To evaluate the six different tools, \BREW is launched six times, each time with the respectively configured underlying \AQL-System.
Each configuration makes the \AQL-System use only one tool.
The setup of a tool is untouched: only required launch parameters are given and the usable memory is specified.
%Thereby the tools ran in their default configuration.
All other options are set to default.

%\begin{table}
%	\caption{Scoring System}
%	\label{tab:scoring_system}
%	\scalebox{0.85}{
%		\begin{tabular}{|c|c|c|}
%			\hline
%			\textbf{From ($\geq$)} & \textbf{To ($<$)} & \textbf{Score} \\
%			\hline
%			0.00 & 0.01 & 0 \\
%			0.01 & 0.25 & 1 \\
%			0.25 & 0.50 & 2 \\
%			0.50 & 0.75 & 3 \\
%			0.75 & 0.90 & 4 \\
%			0.90 & 0.99 & 5 \\
%			0.99 & 1.01 & 6 \\
%			\hline
%		\end{tabular}
%	}
%\end{table}

\textbf{RQ2}. 
As a number of researchers have already carried out a comparison of their own tool with some existing tools, 
we wanted to see what the outcome of a more refined comparison is. 
For the comparison, we chose {\em  F-measure} as our means for evaluating accuracy. 
%To make the result more easily accessible, we furthermore introduced a simple scoring scheme (see Table~\ref{tab:scoring_system}). 
%We divided the interval [0,1] of values for F-measure into seven smaller sub-intervals, each with a different score attached to it. 
%For a whole benchmark set, the scores are summed up. 
For each catergory and tool the average value is computed.
Basically, for comparison the rule ``the larger, the better'' can be applied on the achieved value. 
For evaluation, we again used our refined version of \DroidBench.
\ICCBench is not used since we do not want to intermix benchmark cases.
%A scoring system is introduced to rate the performance of tools in feature restricted areas.
%According to the sum of these scores the probably best tools are named.
%In addition, it is checked which tools are up-to-date wrt. their ability to analyze state-of-the-art apps.
%Clusters of results equal for different tools are highlighted and related to certain features these tools have in common.
%Thereby, we determine strengths and weaknesses in groups of tools that possibly represent best-practices and do-nots for analyses.

\textbf{RQ3}. Regarding scalability, we seek to evaluate whether the tools are able to deal with 
(1) large apps (in terms of code size), (2) a large numbers of apps, (3) ICC and IAC and (4) newer Android versions. 
For (1) we used \DIALDroidBench, a benchmark suite containing 30 large real-world apps. 
Again, we employed \BREW to help us determine the  ground truth for these apps. 
Due to the size of apps, this is anything but straightforward, as a simple manual inspection of all potential flows is not feasible. 
%Considering the large, real world apps collected in DIALDroid-Bench we use our toolchain to partially determine the ground truth for all 30 apps.
We used the following procedure to nonetheless achieve a systematic derivation of a ground truth.
First, we created one positive benchmark case for every pair of sources and sinks.
This resulted in 841,514 potential positive benchmark cases.
Second, we ran all six tools on these cases, ending with a report of 1,007 candidates for privacy leaks.
20 of these potential leaks had been found by two tools.
No leak was found by more than three tools, and there were six benchmark cases for which this occurred.
For all leaks that had been detected multiple times, we manually investigated the source code of the associated apps.
We used the \verb+jadx+~\cite{TL:jadx} decompiler to extract the apps' source code. %\footnote{\url{https://github.com/skylot/jadx}}
To inspect the decompiled code, we then used Android Studio.
The manual inspection led to the confirmation of 22 positive cases.
Four other positive cases had to be rejected, because there exists no data flow between the source (accessing device's location) and the sink (logging e.g. a username).
By means of \BREW these confirmed and rejected benchmark cases have been stored respectively as positive and negative cases in another refined version of \DIALDroidBench.
We are aware that the determined ``ground truth'' is most likely incomplete and does not involve all apps of \DIALDroidBench.  
However, to our knowledge it represents the first available precise (in terms of flows) ground-truth description for a set of real-world apps. 
In cooperation with others~\cite{issta} we plan to extend it in future.

For inspecting the tools' abilities to handle a high number of apps (2), ICC and IAC (3), we used our own intent-matching benchmarks.  
For checking their applicability to newer Android APIs (4), we ran the tools on different versions of our feature-checking test apps compiled and developed with Android Studio as well as on the \DroidBench apps developed with Eclipse.
%as to specifically determine whether the tools are able to correctly find inter-component and inter-app communications.  
% come into play to determine which tools are able to correctly determine possible inter-component and inter-app communications.
%Considering the set of real world apps \BREW is used to determine the most-precise, available ground truth for these apps.

%All tools in our scope are able to analyze at least a subset of these apps.
%How meaningful the results for this subset are is evaluated in detail.
%Even with the help of \BREW~ we cannot determine the ground truth for such apps due to their complexity.
%To this end, we largely overapproximate the ground truth.
%We assume all detectable sources and sinks are connected to each other.
%Then we manually inspect the taint flows found by all 6 tools.
%Whereas all confirmend flows are then marked as positive cases, all others are marked as negative ones.
%Finally, we use \BREW~ again to determine the precision, recall and f-measure for the 6 investigated tools regarding the refined DIALDroid-Bench test suite.

\subsection{Execution Environment}
All experiments were executed on a Debian (Jessie) virtual machine, which has Java~8 (1.8.0\_161) installed.
It was set up to use two cores of an Intel\textregistered\ Xeon\textregistered\ CPU (E5-2695 v3 @ 2.30GHz) and 32~GB memory whereof 30~GB were assigned to the analysis tool as heap space of the respective Java virtual machine.

%% file: content/results.tex
%\eb{Ich denke, wir sollten Section 4 und 5 zusammenziehen.}
%\hw{ich nicht ;-)}
%We next report on our experimental results.
\subsection{RQ1: \RQONE}
\begin{table}[t]
	\caption{Feature Promises}
	\label{tab:promises_features}
	\resizebox{\columnwidth}{!}{
		\begin{tabular}{|c|c|c|c|c|c|c|c|c|c|c|c|c|c|c|c|c|}
			\hline
			\raisebox{2.15cm}{~} &
			\raisebox{0.6cm}{\multirow{2}{*}{\rotatebox{90}{\textbf{Aliasing}~}}} &
			\raisebox{0.2cm}{\multirow{2}{*}{\rotatebox{90}{\textbf{Static}~}}} &
			\raisebox{0.8cm}{\multirow{2}{*}{\rotatebox{90}{\textbf{Callbacks}~}}} &
			\raisebox{0.7cm}{\multirow{2}{*}{\rotatebox{90}{\textbf{Lifecycle}~}}} &
			\raisebox{1.8cm}{\multirow{2}{*}{\rotatebox{90}{\textbf{Inter-Procedural}~}}} &
			\raisebox{1.0cm}{\multirow{2}{*}{\rotatebox{90}{\textbf{Inter-Class}~}}} &
			\raisebox{0cm}{\multirow{2}{*}{\rotatebox{90}{\textbf{IAC}~}}} &
			\raisebox{1.05cm}{\multirow{2}{*}{\rotatebox{90}{\textbf{ICC \footnotesize(explicit)}~}}} &
			\raisebox{1.1cm}{\multirow{2}{*}{\rotatebox{90}{\textbf{ICC \footnotesize(implicit)}~}}} &
			\rotatebox{90}{\textbf{Flow-}~} &
			\rotatebox{90}{\textbf{Context-}~} &
			\rotatebox{90}{\textbf{Field-}~} &
			\rotatebox{90}{\textbf{Object-}~} &
			\rotatebox{90}{\textbf{Path-}~} &
			\raisebox{1.7cm}{\multirow{2}{*}{\rotatebox{90}{\small \textbf{ThreadAwareness}~}}} &
			\raisebox{0.9cm}{\multirow{2}{*}{\rotatebox{90}{\textbf{Reflection}~}}} \\
			\textbf{Tool} & & & & & & & & & & \multicolumn{5}{c|}{\textbf{Sensitivity}} & &	\\
			\hline
			\Amandroid	& \textcircled{$+$} & \textcircled{$\star$} & \textcircled{$\star$} & \textcircled{$\star$} & \textcircled{$\star$} & \textcircled{$\star$} & \textcircled{$\star$} & \textcircled{$+$} & \textcircled{$+$} & {\color{red}\textcircled{$-$}} & \textcircled{$\star$} & \textcircled{$+$} & \textcircled{$+$} & $-$ & $\star$ & $+$ \\
			\hline
			\DIALDroid	& & & & & & & {\color{red}\textcircled{$-$}} & \textcircled{$\star$} & {\color{red}\textcircled{$-$}}$^\dagger$ & & & & & & & \\
			\hline
			\DidFail		& \textcircled{} & \textcircled{} & \textcircled{} & \textcircled{} & \textcircled{} & \textcircled{} & \textcircled{} & \textcircled{} & \textcircled{} & \textcircled{} & \textcircled{} & \textcircled{} & \textcircled{} & & & \\
			\hline
			\DroidSafe	& \textcircled{$+$} & \textcircled{$\star$} & {\color{red}\textcircled{$-$}} & \textcircled{$\star$} & \textcircled{$\star$} & \textcircled{$\star$} & \textcircled{$+$} & \textcircled{$+$} & \textcircled{$+$} & $-$ & {\color{red}\textcircled{$-$}} & \textcircled{$+$} & \textcircled{$+$} & $-$ & $\star$ & $+$ \\
			\hline
			\FlowDroid	& \textcircled{$+$} & \textcircled{$\star$} & \textcircled{$\star$} & \textcircled{$\star$} & \textcircled{$\star$} & \textcircled{$\star$} & $-$ & $-$ & $-$ & \textcircled{$\star$} & \textcircled{$\star$} & \textcircled{$+$} & \textcircled{$\star$} & $-$ & $\star$ & $-$ \\
			\hline
			\IccTA		& \textcircled{$+$} & \textcircled{$\star$} & \textcircled{$\star$} & \textcircled{$\star$} & \textcircled{$\star$} & \textcircled{$\star$} & \textcircled{$\star$} & \textcircled{$\star$} & \textcircled{$+$} & \textcircled{$\star$} & \textcircled{$\star$} & \textcircled{$+$} & \textcircled{$\star$} & $-$ & $\star$ & $-$ \\
			\hline
			\multicolumn{17}{c}{\small \textcircled{~} supported, $\star$ confirmed, $+$ partially confirmed, $-$ not confirmed, $^\dagger$ aborted} \\
			%\multicolumn{17}{c}{$^\dagger$ aborted, $^*$ not meaningful}
		\end{tabular}
	}%
\end{table}

\paragraph{Feature promises.}
Each tool promises to support a certain set of features (see original paper and summary in surveys~\cite{DBLP:journals/infsof/LiBPRBOKT17, DBLP:journals/tse/SadeghiBGM17}).
On the basis of our feature-checking benchmark cases, we verified which features are supported.
Table~\ref{tab:promises_features} summarizes the results.
Each row stands for one tool; each column represents one feature. The entries describe the promises and their degree of fulfillment: 
the symbol $\star$ stands for {\em full} support, $+$ for {\em partial} support  and 
$-$ for all feature-checking benchmark cases having been failed. 
Furthermore, if the symbol is circled (\textcircled{~}), the corresponding feature was promised to be supported.
Consequently, a promise violation is denoted as a circled minus symbol ({\color{red}\textcircled{$-$}}).
In the table, five promise violations appear. We shortly describe the reasons for these. 
\begin{enumerate}
\item \Amandroid failed to detect the correct order of statements causing it to falsely determine a taint flow.
\item 
\DIALDroid struggled in dealing with implicit intents for ICC and IAC.
In case of the test app for implicit ICC, its execution was aborted at the end due to a database error.
%Thus the correct result possibly has been found.
\item \DroidSafe failed to handle callbacks correctly.  
This is surprising, as the paper claims that due to  its flow-insensitivity it can handle callbacks more easily. In addition \DroidSafe seems to over-approximate in case of the context-sensitivity check, violating a second promise.
However, its development has stopped in 2016, and the last supported Android version is 4.4.1.
Hence, it is plausible that some features are no longer supported.
\end{enumerate}  
For \DidFail, we could not check whether it keeps its promises since it cannot handle newer apps (see Table~\ref{tab:up_to_date}), nor apps that do not name its targeted Android API version in its manifest.
To do so has become optional along with the establishment of Android Studio as dedicated development platform in 2014.

In summary, according to our feature-checking benchmarks, all promises except five are kept and most of the promised features are fully supported.

\begin{figure}
	\centering
	\includegraphics[trim={1.5cm 3.1cm 6.7cm 2cm},clip,width=\columnwidth]{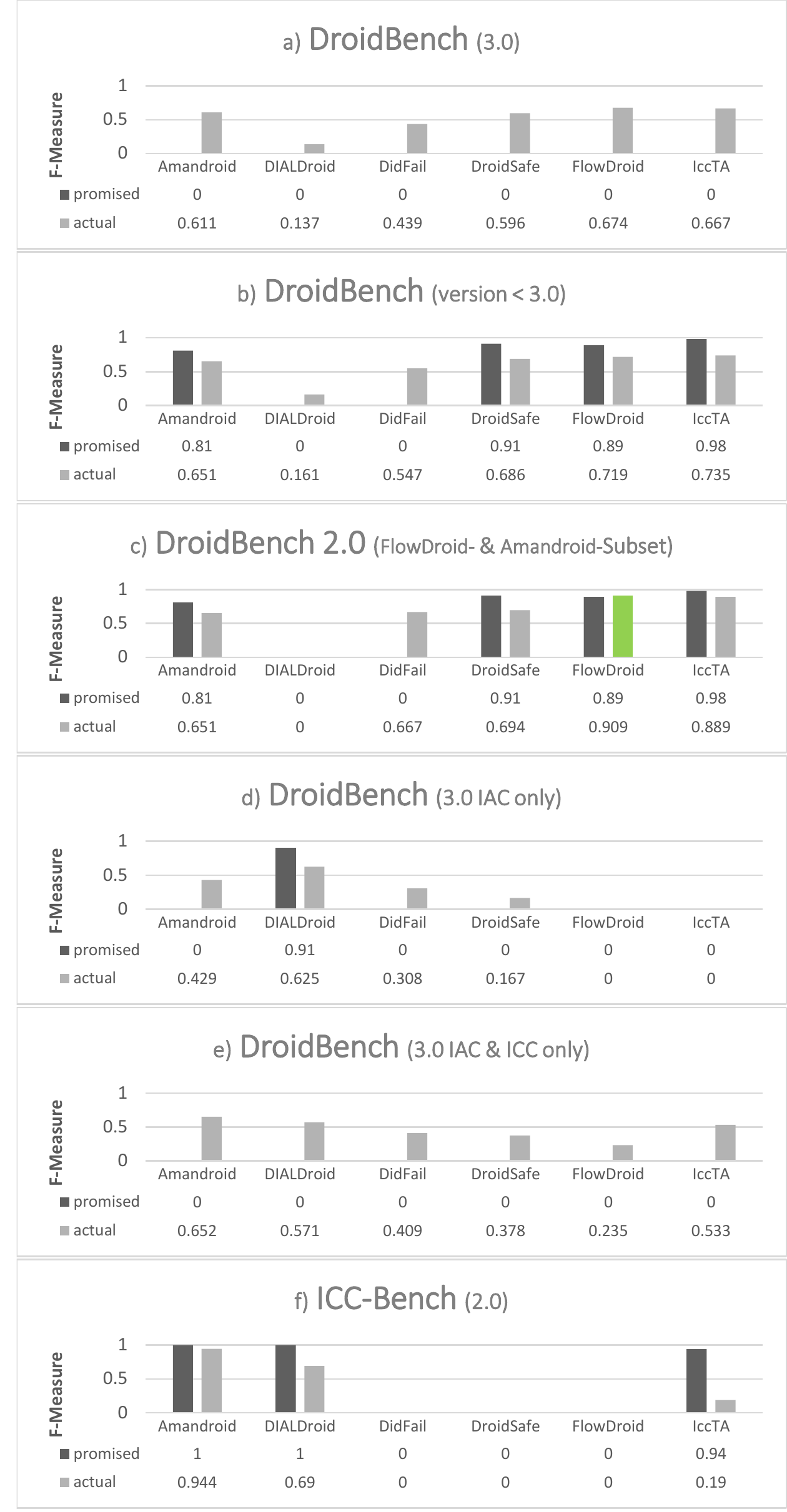}
	\caption{Accuracy Promises}
	\label{fig:promises_prfm}
\end{figure}

\paragraph{Accuracy promises.}  
Accuracy is typically evaluated by the metrics precision, recall and F-measure -- the harmonic mean of the first two.
For the sake of clarity, and because it best represents the overall accuracy, we only report the F-measure here.
Figure~\ref{fig:promises_prfm} depicts the F-measures for all six tools on different sets of micro-benchmark apps. 
We used different sets because the promises themselves typically refer to different benchmark suites. 
The dark bar always represents the {\em promised} value, the brighter one the {\em actual} value determined in our experiments. 

\textbf{\DroidBench version 3.0}: Figure~\ref{fig:promises_prfm}\textit{a} shows the F-measure values for the current version (3.0) of \DroidBench.
Since no paper promised anything for the complete 3.0 set, there are no dark (promised) bars shown.
All tools have an accuracy of about 60\% apart from \DIALDroid and \DidFail which have less.
60\% does not sound to be an inspiring confidence, but a lot of distinct features are exploited in \DroidBench 3.0, specifically such features designed to challenge existing tools.
Thus it was to be expected that each tool makes mistakes at some point.
Furthermore, we find that \DIALDroid has a very low value in Figure~\ref{fig:promises_prfm}\textit{a} as well as in most of the others. 
This is because \DIALDroid is designed for ICC and IAC cases only (tracking no intra-component flows), and consequently fails in all other cases.
\DidFail only reached an F-measure of $0.439$ for \DroidBench 3.0. This is because \DidFail is the oldest and fewest updated tool considered.
%No other tool should and is not worse \hw{doppelte Verneinung: no other .. is not worse?} than \DidFail for any benchmark set.

\textbf{\DroidBench version < 3.0}: Most tools were proposed when \DroidBench 3.0 did not exist, hence an older version of \DroidBench was used.
The bar chart in Figure~\ref{fig:promises_prfm}\textit{b} shows the promised and actual values for \DroidBench before version 3.0.
On this, no tool achieved its promised value.
With a relative deviation of 9\% \FlowDroid is closest to its promise.
The other tools are at about 19\% to 25\% below the promised value. % Amandroid:20, DroidSafe:25, FlowDroid: 19
All tools nevertheless achieved better values for this set than for the 3.0 set.

\textbf{\DroidBench version 2.0 (subset)}: The results improve if we only take a certain subset of \DroidBench 2.0 into account (see Figure~\ref{fig:promises_prfm}\textit{c}).
This subset has been used in the papers proposing \FlowDroid and \Amandroid.
However, only \FlowDroid is able to keep its promise for this subset.

\textbf{\DroidBench 3.0 IAC only}: Bosu et al.~\cite{DBLP:conf/ccs/BosuLYW17} evaluated \DIALDroid for all IAC cases of \DroidBench 3.0 
and claimed to achieve an F-measure of $0.91$ (see Figure~\ref{fig:promises_prfm}\textit{d}).
In our experiments we could only reproduce an F-measure of $0.625$ for the same subset.
Nonetheless, this is the best value for this subset.
All other tools could not even reach 50\%, which makes them inappropriate.
As expected \FlowDroid was not able to detect any inter-app flows.
Unfortunately, \IccTA was also unable to handle those although it should be 
as claimed in the associated paper and determined by the feature-checking benchmark.
A closer look at the individual results reveals that \IccTA could not resolve flows between \texttt{setResult(..)} and \texttt{onActivityResult(..)} statements.
The intra-app parts of the leak, however, could  be found as well  as the connection between the two apps involved.
It seems to be a tiny but decisive bug in \IccTA's implementation.

\textbf{\DroidBench 3.0 IAC and ICC only}: Once we add all ICC cases of \DroidBench, all tools except \DIALDroid become more accurate on average (see Figure~\ref{fig:promises_prfm}\textit{e}).
\IccTA improves the most and achieves an F-measure of $0.533$.
The best value is achieved by \Amandroid which overtook \DIALDroid for this extended subset.
\FlowDroid's value is not $0$ because some of DroidBench's ICC cases communicate values through static fields or use statements receiving or sending intents as only sources or sinks. \FlowDroid can handle both.

\textbf{\ICCBench}: Finally, we inspected \ICCBench (see Figure~\ref{fig:promises_prfm}\textit{f}).
For this micro-benchmark set we have only three promises.
Whereas \Amandroid almost keeps its promise, \DIALDroid lacks 31\%.
\IccTA underperforms the most, because of the reason discussed above.

To conclude, we could not reproduce the accuracy that was claimed in the proposing papers apart from one promise made by 
Arzt et al.~\cite{DBLP:conf/pldi/ArztRFBBKTOM14} for \FlowDroid considering a small set of benchmark cases (see Figure~\ref{fig:promises_prfm}\textit{c}).
In consequence, the answer to RQ1 is thus that the tools in general keep only parts of their promises.  

% -------------------------------------------------------------
\subsection{RQ2: \RQTWO}

\begin{table}
	\caption{F-Measure Scores}
	\label{tab:scores}
	\resizebox{\columnwidth}{!}{
		\begin{tabular}{|r|l|c|c|c|c|c|c|c|}
			\hline
			ID & Category & \rotatebox{90}{\textbf{\DIALDroid~}} & \rotatebox{90}{\textbf{\DidFail~}} 	& \rotatebox{90}{\textbf{\DroidSafe~}} & \rotatebox{90}{\textbf{\IccTA~}} & \rotatebox{90}{\textbf{\FlowDroid~}} & \rotatebox{90}{\textbf{\Amandroid~}} & \O \\
			\hline
			1  & FieldAndObjectSensitivity	& 0.000	& \cellcolor{OliveGreen!30}0.800	& \cellcolor{OliveGreen!20}0.667	& \cellcolor{OliveGreen!50}1.000	& \cellcolor{OliveGreen!50}1.000	& \cellcolor{OliveGreen!50}1.000				& 0.745 \\
			2  & Callbacks					& 0.000	& \cellcolor{OliveGreen!30}0.769	& \cellcolor{OliveGreen!20}0.667	& \cellcolor{OliveGreen!30}0.897	& \cellcolor{OliveGreen!30}0.897	& \cellcolor{OliveGreen!20}0.500				& 0.622 \\
			3  & UnreachableCode			& 0.000	& \cellcolor{OliveGreen!50}1.000	& 0.000	& \cellcolor{OliveGreen!30}0.857	& \cellcolor{OliveGreen!50}1.000	& \cellcolor{OliveGreen!30}0.857											& 0.619 \\
			4  & AndroidSpecific			& 0.000	& \cellcolor{OliveGreen!10}0.429	& \cellcolor{OliveGreen!40}0.900	& \cellcolor{OliveGreen!30}0.842	& \cellcolor{OliveGreen!40}0.900	& \cellcolor{OliveGreen!20}0.625				& 0.616 \\
			5  & GeneralJava				& 0.000	& \cellcolor{OliveGreen!20}0.611	& \cellcolor{OliveGreen!30}0.780	& \cellcolor{OliveGreen!30}0.762	& \cellcolor{OliveGreen!30}0.810	& \cellcolor{OliveGreen!20}0.703				& 0.611 \\
			6  & EmulatorDetection			& 0.000	& 0.000	& \cellcolor{OliveGreen!20}0.500	& \cellcolor{OliveGreen!40}0.966	& \cellcolor{OliveGreen!40}0.966	& \cellcolor{OliveGreen!40}0.966											& 0.566 \\
			7  & Lifecycle					& 0.000	& \cellcolor{OliveGreen!10}0.400	& \cellcolor{OliveGreen!40}0.933	& \cellcolor{OliveGreen!20}0.737	& \cellcolor{OliveGreen!30}0.769	& \cellcolor{OliveGreen!20}0.545				& 0.564 \\
			8  & {\small InterComponentCommunication}	& \cellcolor{OliveGreen!20}0.538	& \cellcolor{OliveGreen!10}0.452	& \cellcolor{OliveGreen!10}0.480	& \cellcolor{OliveGreen!20}0.706	& 0.348	& \cellcolor{OliveGreen!30}0.750	& 0.546 \\
			9  & Threading					& 0.000	& \cellcolor{OliveGreen!20}0.667	& 0.000	& \cellcolor{OliveGreen!20}0.667	& \cellcolor{OliveGreen!50}1.000	& \cellcolor{OliveGreen!20}0.667											& 0.500 \\
			10 & ArraysAndLists				& 0.000	& \cellcolor{OliveGreen!10}0.444	& \cellcolor{OliveGreen!20}0.667	& \cellcolor{OliveGreen!20}0.500	& \cellcolor{OliveGreen!20}0.615	& \cellcolor{OliveGreen!20}0.545				& 0.462 \\
			11 & Aliasing					& 0.000	& 0.000	& 0.000	& \cellcolor{OliveGreen!20}0.667	&\cellcolor{OliveGreen!20} 0.667	& \cellcolor{OliveGreen!20}0.500																		& 0.306 \\
			12 & InterAppCommunication		& \cellcolor{OliveGreen!20}0.625	& 0.308	& 0.167	& 0.000	& 0.000	& \cellcolor{OliveGreen!10}0.429																									& 0.255 \\
			13 & Reflection					& 0.000	& 0.095	& 0.333	& 0.095	& 0.095	& 0.182																																							& 0.133 \\
			14 & DynamicLoading				& 0.000	& 0.000	& 0.000	& 0.000	& 0.000	& \cellcolor{OliveGreen!20}0.500																																& 0.083 \\
			15 & Native						& 0.000	& 0.000	& 0.000	& 0.000	& 0.000	& 0.333																																							& 0.056 \\
			16 & ImplicitFlows				& 0.000	& 0.000	& 0.000	& 0.000	& 0.000	& 0.000																																							& 0.000 \\
			17 & Reflection\_ICC			& 0.000	& 0.000	& 0.000	& 0.000	& 0.000	& 0.000																																							& 0.000 \\
			18 & SelfModification			& 0.000	& 0.000	& 0.000	& 0.000	& 0.000	& 0.000																																							& 0.000 \\
			\hline
			   & \O							& 0.065	& 0.332	& 0.339	& 0.483	& 0.504	& 0.506	& \\
			\hline
		\end{tabular}
	}
\end{table}

For the comparison of tools wrt.~accuracy, we used \DroidBench 3.0 (and its specific categories).
% and the scoring scheme explained in Section~\ref{sec:evaluation}. 
%\DroidBench is divided into different categories.
%All apps mainly exploit category specific functionalities in order to create or hide privacy leaks.
Table~\ref{tab:scores} shows all categories of \DroidBench 3.0 in its second column.
The following six columns show the F-measure of each tool for all benchmark cases in the associated category.
%The scores are based on the f-measure each tool achieved for all benchmark cases of the selected area.
%In Table~\ref{tab:scoring_system} the scoring system is listed.
%For 7 intervals scores from 0 to 6 can be reached.
%Unless a tool was unable to find anything the score will at least be $\geq 1$.
%Then every $0.25$ step, one more point can be scored.
%The top f-measure values of more than $0.75$ or $0.90$ and the best ($1.00$) are respectively rated with a score of $4$, $5$ and $6$.
%All scores for all categories are denoted as explained in Table~\ref{tab:scores}.
The categories supported best are placed at the top of the table.
Additionally, a color scheme has been added to emphasize each tool's performance:
the darker the background of a cell is, the higher the F-measure.

We find that the first 11 of 18 categories are handled properly by most tools.
The F-measure values achieved in Category~12, namely InterAppCommunication, are inadequate.
According to the promises made, the tools should be able to analyze inter-app cases.
In particular, \DIALDroid should achieve a top value here but it does not excel at an F-measure of $0.625$.
The remaining six categories are insufficiently handled by all tools apart from some special cases.
However, this matches our expectations since no tool claimed to be able to handle such cases. 

Before we conclude this section,
a few remarks regarding the runtime: while most tools needed on average 25 seconds to analyze one app, \DroidSafe required more than 200 and \FlowDroid less than 10 seconds on average.
In addition, \DroidSafe timed out in 25 cases by exceeding a maximal execution time of 10 minutes.
This also happened in case of \DidFail and \IccTA but only once.

In summary, we see that there is no single ``best'' tool. 
Every tool has at least one other tool performing better in at least one category. 
Overall (see sums in the final row), \Amandroid ($0.506$) and \FlowDroid ($0.504$) score best wrt.~accuracy.

%\begin{figure}
%	\centering
%	\includegraphics[width=0.85\columnwidth]{figures/scores}
%	\caption{Scores visualized}
%	\label{fig:scores}
%\end{figure}

% -------------------------------------------------------------
\subsection{RQ3: \RQTHREE}

\begin{table}
	\caption{DIALDroid-Bench results}
	\label{tab:rwa}
	\resizebox{\columnwidth}{!}{
		\begin{tabular}{|c|c|c|c|}
			\hline
			%                          No timeout (30 mins)
			\multirow{2}{*}{Tool}	& {\small Number of successfully}	& {\small Analysis-time}		& {\small Number of true} \\
									& {\small analyzed apps}			& {\small per app (minutes)}	& {\small / false positives} \\
			\hline
			\Amandroid	& 21	& 8		& ~1\quad/\quad0 \\
			\DIALDroid	& 20	& 10	& ~0\quad/\quad4 \\
			\DidFail	& 27	& 9		& 21\quad/\quad4 \\
			\DroidSafe	&  2	& 5		& ~0\quad/\quad0 \\
			\FlowDroid	& 18	& 2		& 22\quad/\quad0 \\
			\IccTA		& 18	& 4		& ~6\quad/\quad0 \\
			\hline
		\end{tabular}
	}
\end{table}
Table~\ref{tab:rwa} (second column) shows how many real-world apps each tool was able to analyze without exceeding the maximal execution time of 30 minutes.
The third column shows the average execution time of each tool, 
the last column how many of expected results could be matched by each tool.
There is no tool that was able to analyze all apps, and apart from \FlowDroid every tool on average required 
more than two minutes to analyze an app.
%\begin{wraptable}[12]{l}{4.75cm}
The results for the newly defined ground truth (last column) reveal that \FlowDroid currently seems to be the best choice to reliably deal with large apps, 
since all other tools missed some confirmed leaks or falsely detected rejected ones.
In addition, in case of \Amandroid, \DIALDroid, \DroidSafe and \IccTA most of the 22 confirmed leaks remain undetected.

\begin{figure}[h]
	\centering
	\includegraphics[trim={1.5cm 23.6cm 6.7cm 2cm},clip,width=\columnwidth]{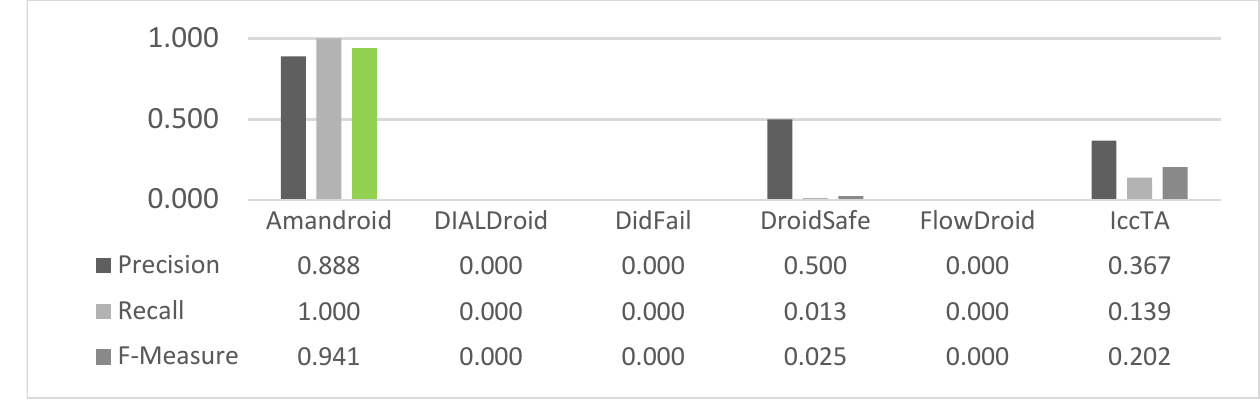}
	\caption{Intent-Matching: Precision, Recall, F-Measure}
	\label{fig:intent_matching}
\end{figure}

%\begin{table}[h]
\begin{wraptable}[10]{l}{4.5cm}
	\vspace{-0.2cm}
	\caption{Up-to-date Status}
	\label{tab:up_to_date}
	\scalebox{0.75}{
		\begin{tabular}{|c|c|c|c|}
			\hline
			\multirow{2}{*}{\textbf{Tool}} & \textbf{Eclipse} & \multicolumn{2}{c|}{\textbf{AndroidStudio}} \\
			& \textbf{API $\leq$ 19} & \textbf{API 19} & \textbf{API 26} \\
			\hline
			\Amandroid	& \ding{51} & \ding{51} & $-$ \\
			\DIALDroid	& \ding{51} & \ding{51} & \ding{51} \\
			\DidFail	& \ding{51} & $-^\dagger$		& $-^\dagger$ \\
			\DroidSafe	& \ding{51} & \ding{51} & $-^\dagger$ \\
			\FlowDroid	& \ding{51} & \ding{51} & \ding{51} \\
			\IccTA		& \ding{51} & \ding{51} & $-^\dagger$ \\
			\hline
			\ApkCombiner	& \ding{51} & \ding{51} & $-^\dagger$ \\
			\hline
			\multicolumn{4}{c}{\ding{51} supported, $-$ fails, $^\dagger$ crashes} \\
		\end{tabular}
	}
\end{wraptable}

The intent-matching benchmark results further support this conclusion (see Figure~\ref{fig:intent_matching}).
Apart from \Amandroid, no tool is able to accurately match the action, category or data field.
%Therefor Amandroid's result is surprisingly good with an f-measure value of $0.941$.

%Nevertheless, according to our scoring system two tools are tied for the position of the best tool, namely Amandroid and FlowDroid.
%These two most cited\footnote{According to Google Scholar FlowDroid was cited 800 times and Amandroid 217 times followed by IccTA with 207 cites.} tools both achieve a score of 48.
%As a tiebreaker we check whether both tools are able to analyze state-of-the-art apps built for newer Android versions released after June 2014\footnote{In June 2014 Android 4.4.4 the last Android version using API 19 was released.}.
%Both tools are frequently updated and have been updated in 2017, thus both are expected to be able to deal with newer Android versions.
%Surprisingly, Amandroid can not (see Table~\ref{tab:up_to_date}).
%In contrast to other tools in our scope, Amandroid fails its analysis without crashing.

\noindent
Finally, we investigated whether the tools are able to handle newer Android versions.
Table~\ref{tab:up_to_date} shows the outcomes. 
\Amandroid fails to perform a proper analysis, however without crashing. 
\DroidSafe, \IccTA and \ApkCombiner all crash while analyzing apps built for an API above 19, which is supported by the majority (82.3\%) of Android devices.\footnote{\url{https://developer.android.com/about/dashboards} (01/03/2018)}
A common cause is a tool-dependency on the \verb+Apktool+~\cite{TL:ApkTool}.
%\footnote{\url{https://ibotpeaches.github.io/Apktool/}}
Old versions of it fail to decompile newer Android apps.
The same happens to the \ApkCombiner.
Thereby \Amandroid, \DroidSafe, \FlowDroid and \IccTA lose their ability to analyze inter-app scenarios.
%Regarding RQ2, the best tool in this context should be able to handle state-of-the-art apps.
%Consequently FlowDroid, although it cannot handle ICC and IAC cases, currently represents the best tool among the 6 tools in our scope.
As discussed before, \DidFail fails to analyze newer apps developed with Android Studio.
In summary,  the ans\-wer to RQ3 is: each tool in our scope still has shortcomings when it comes to analyzing real-world apps.

\subsection{Threats to Validity}
The main threat to the validity of our experiments arises from the manual, though tool-assisted, 
definition of the expected results.
However, currently we see no way around it because the tools that could potentially be used to derive the ground truth are 
at the same time the tools we want to evaluate.
Thus we cannot rely on a single tool to generate the ground truth. 
Moreover, \tool allows us to refine the expected result definitions multiple times and thereby to achieve precise results. 
Similarly, also other experts can use \tool to define and refine their own benchmarks. 
%Thereby, an expected result definition will become a valid ground truth over time.

In all our experiments the tools have been executed in their default configuration.
Only the available memory has been changed according to our system's setup.
Some tools may produce different results when executed with specific launch parameters.
These results may be more accurate or less, computed faster or slower, and might thus change the outcome of our experiments. 
For example, \FlowDroid has an option to activate the tracking of implicit flows.
Consequently, its F-measure value would have been greater than $0$ in category 15 (see Table~\ref{tab:scores}).
We restricted our experiments to the default configuration nonetheless because this is the one which a non-expert software developer is likely going to use.

%However, the most used configuration most likely is the default one.
%Hence, we restricted our evaluation to this configuration.

\todo[inline]{Camera-ready}{Added now}
Another threat to validity are the metrics precision, recall and F-measure that we and others frequently used to measure for accuracy. 
For some feature-restricted parts of e.g.~\DroidBench, these metrics are misleading.
The Aliasing subset, for example, comprises four benchmark cases, three of which are negative cases.
A tool capable of correctly answering these three negative cases will still have a precision, recall and F-measure value of 0 if it just fails the single positive case.

Finally, our implementation of the  \AQL-System may contain bugs.
In particular, the converter used to translate tool-specific answers into \AQL-Answers must work as intended in order to produce correct and meaningful results.
We extensively tested the converter and fixed all errors. 
Due to the imprecise format of some tools' results, sources or sinks are sometimes (more specifically, for \DidFail) not uniquely identifiable while converting it into the \AQL format.
Therefore, the converter over-approximates, i.e. all candidates are taken into account as source or sink respectively.
Considering our experiments on real-world apps, \BREW similarly over-approximated during the identification of sources and sinks in case of \Amandroid, \DidFail and \DroidSafe.
Thus, method calls are matched by method names without considering the parameters given as input.
%Due to the mismatch between \AQL-Answers and tool results (e.g.~concerning source identification), the converter sometimes (more specifically, for \DidFail) needs to over-approximate results in order to fairly treat tools. 
Such aspects have already been taken into consideration. 
%During our experimentation  all converter related errors have been fixed.
%Hence, the converter created for the 6 tools in our scope should work as expected.
%In particular, the converter developed for \DidFail overapproximates some results since a source cannot be uniquely identified if, for example, two function calls of the same type are found inside the same method.
%Such problems can be eliminated and converters can become superfluous once a tool directly produces a machine readable result like \AQL-Answers (XML).

%% file: content/relatedwork.tex
Providing an overview of analysis tools for Android apps is the topic of three recent surveys~\cite{DBLP:journals/infsof/LiBPRBOKT17, DBLP:journals/tse/SadeghiBGM17, DBLP:journals/csur/ReavesBGABCDHKS16}.
These works collect and summarize tools and their functionality as outlined in research papers. 
They provide no systematic experimentation to assess and compare tools, in particular not with respect to their promises. 

A thorough comparison of Android analysis tools has so far been difficult due to the lack of precisely defined benchmarks. 
This situation is different in other (analysis) contexts. 
Competitions like the annual Competition on Software Verification (SV-COMP~\cite{DBLP:conf/tacas/Beyer17}), 
the SAT-solving competition\footnote{\url{http://satcompetition.org}} 
or the Hardware Model Checking Competition (HMCC~\cite{DBLP:conf/fmcad/BiereDH17}) 
provide well-defined benchmarks in different categories with precisely fixed outcomes. 
Often, they do not just require participating tools to give yes/no answers, 
but in addition to provide witnesses or proofs of their results.  
With the \AQL, we already have a format for witnesses of taint flows available. 
%Witnesses should be independently checkable. 
%Such competitions often boost tool development.

Finally, there are more tools which potentially fit in our scope. 
{\sc Apposcopy}~\cite{DBLP:conf/sigsoft/FengDAA14}, {\sc WeChecker}~\cite{DBLP:conf/wisec/CuiWHXZY15} and {\sc Separ}~\cite{DBLP:conf/dsn/BagheriSBM16} should fit perfectly,  
however are not publicly available.
SCanDroid~\cite{scandroid} is publicly available and fits into our scope as well, nonetheless it is largely outdated and cannot produce results for any considered (micro) benchmark.
{\sc DroidInfer}~\cite{DBLP:conf/issta/HuangDMD15} employs an interesting type-based approach.
However, in this particular case it requires a lot of effort to build a converter to extract the determined taint flows, because of its uncommon result structure.
Additionally, the tool seemed not ready for competitive comparison since its execution fails for most micro-benchmark cases.
Thus, we decided to omit the tool.
Other available tools do not fit into our scope due to their result representation even though they inspect privacy leaks.
For example, {\sc HornDroid}~\cite{DBLP:conf/eurosp/CalzavaraGM16} determines sinks and provably shows that these can be reached by taint flows.  
It fails, however, to name sources, which is why we cannot determine which specific flow is found.

%% file: content/conclusion.tex
In this paper, we reported on the results of a reproducibility study on static taint analysis tools for Android apps. 
To support our own as well as similar studies, we developed a framework for inferring data leaks in test apps 
and for automatically running tools on benchmark sets. 
With the help of this framework, we assembled precise benchmark suites and re-evaluated six existing tools on them. 
In the evaluation, we in particular studied the handling of specific features, the accuracy of tools and their relation to the promised values. 
The results indicate that studies and benchmarks like ours are indeed needed to provide a solid ground for a fair and unbiased comparison of tools.